# Validation and demonstration of the AEFC as a practical safeguards tool for inventory verification

G. Long[a], G. Gabrieli[b], I. Levy[c], A. Pesach[c], I. Zilberman[b], K. Ben-Meir[b], O. Ozeri[b], L. Zilberfarb[b], O. Rivin[c], A. Krakovich[b], I. Martinez[a], C. Rael[a], M. Watson[a], A. Trahan[a], M. L. Ruch[a], U. Steinitz[b]

Affiliations:

[a] Los Alamos National Laboratory

[b] Soreq Nuclear Research Centre

[c] Nuclear Research Centre – Negev

Abstract:

The Advanced Experimental Fuel Counter (AEFC) is a nondestructive assay (NDA) instrument designed to determine the residual fissile mass in irradiated fuel assemblies for safeguards verification purposes. This is done by actively interrogating an assembly with a neutron source and measuring the total (Singles) and correlated (Doubles) neutron count rates resulting from induced fissions in the irradiated nuclear fuel and relating those rates to the residual fissile mass using calibration curves. Comprehensive NDA measurements of the irradiated fuel inventory at Israeli Research Reactor 1 (IRR-1) were taken with the AEFC to validate a set of previously developed calibration curves. During the campaign, measurements were acquired of 32 standard fuel assemblies and three control assemblies in just nine days. This is a significant majority of the research reactor's irradiated fuel inventory and the largest data set gathered by the AEFC to date.

Many of the fuel assemblies measured during the campaign had much shorter cooling times than those assemblies previously measured with the instrument. Calibration curves developed from previous AEFC deployments were used to determine the residual $^{235}$U mass in the measured standard fuel assemblies. The results of the campaign demonstrated that the AEFC can be used to estimate the $^{235}$U mass remaining in a large number of irradiated fuel assemblies with 1%-5% uncertainty in a reasonable amount of time despite operating in a high dose environment.

# 1. Introduction

The Advanced Experimental Fuel Counter (AEFC) is a nondestructive assay instrument for characterizing spent research reactor fuel [1]. The instrument actively interrogates fuel elements with a $^{252}$Cf neutron source and six $^3$He tubes for detecting the resulting induced-fission neutrons. The AEFC performs active neutron coincidence counting to quantify the fissile mass remaining in fuel assemblies (FAs) after burning in a reactor and is intended for safeguards inspections. Equipped with an ion chamber, the device also obtains supplemental gamma dose and axial profile data. The system is typically set-up at the bottom of a spent fuel pool for underwater fuel measurements. As a safeguards tool, it is designed to achieve acceptable measurement uncertainty while maintaining reliability and operational ease of use.

In 2018, we conducted a measurement campaign at the Israeli Research Reactor 1 (IRR-1), which is a 5 MW open pool reactor using highly enriched uranium (HEU) materials test reactor (MTR) fuel [2]. In that campaign, the $^{235}$U content of 14 FAs was estimated using the AEFC and compared to other measurements and calculations, generating calibration curves for the AEFC. A campaign was then undertaken in 2022 [3] at IRR-1 to validate the calibrations that were developed in the first campaign by diversifying the fuel burnup level in the assayed FAs and to extend the operating conditions of the AEFC to FAs with shorter cooling times, thus exposing the AEFC to higher gamma doses. In this second campaign, we also measured a nearly complete inventory tally of a reactor's standard FAs for the first time.

Here, we report the main findings from the comprehensive campaigns, discuss sources of uncertainty with emphasis on gamma-ray interference, analyze calibration methods, and draw conclusions for further improvements. We focus on the AEFC's robustness and accuracy under different measurement conditions, specifically, higher gamma dose rate, which is inherent to burnt FAs. This work will allow us to reevaluate AEFC measurement uncertainties and to consider additional calibration schemes.

# 2. Materials and Methods

## 2.1 Measurement Overview

The first AEFC deployment at the IRR-1 took place in January 2018. During that campaign, we used the AEFC to measure 14 standard FAs (denoted "FS") and 1 control assembly ("FC"). Two calibration curves covering a wide burnup range were created from these measurements: one using total neutron count rate (denoted "Singles rate") and one with the correlated coincidence neutron count rate ("Doubles rate") [4]. Both calibration curves were validated with three methods: rhenium gamma transmission measurements (RGT) [5], gamma emission measurements of the FAs, and calculations of more than 30 years of burnup history [6]. The validated calibration curves allow the use of the AEFC as an independent tool for measuring residual fissile mass in an FA.

During the 2018 campaign, counting statistics was the dominant source of uncertainty for the Doubles-rate-based fissile mass estimate, by a considerable margin [2]. Therefore, for the 2022 campaign, we used a 3.5 times stronger $^{252}$Cf interrogation source, increasing the induced fission rate and reducing statistical uncertainties in the Singles and Doubles count rates. We also prolonged the individual measurements' acquisition time from 5 minutes in 2018 to 8 minutes in the 2022 campaign to further reduce counting statistics uncertainty in the Doubles rate.

At the 2022 campaign, the signals from the AEFC's six $^3$He tube preamplifiers were combined using an OR box before being fed into the JSR-15 [7] shift register connected to a laptop running INCC 5.1.2 [8]. The AEFC's ion chamber was powered by an external high voltage (HV) supply and connected to a current-to-pulse converter whose output was connected to the JSR-15. Figure 1 shows images of the instrument in the IRR-1 pool (left) and of the acquisition electronics (right).

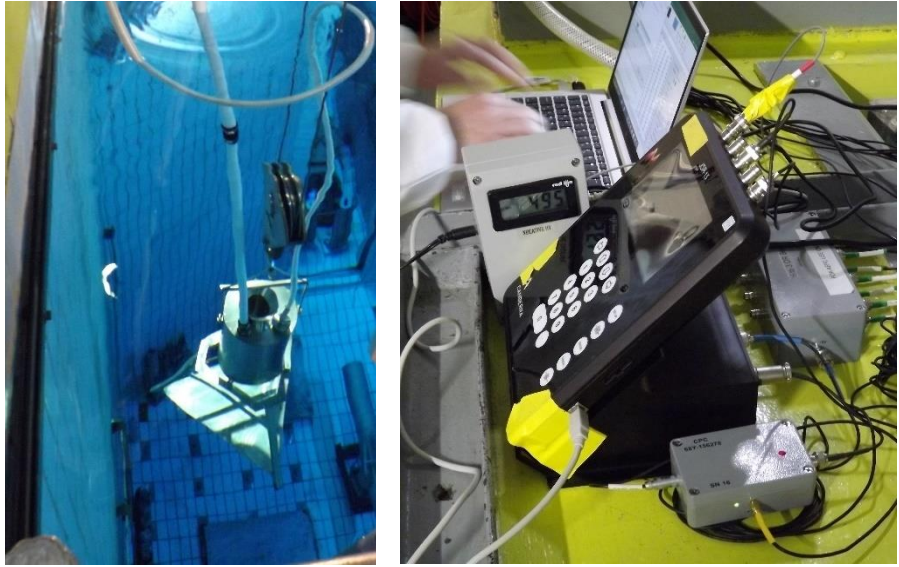

*Figure 1. AEFC lowered to the bottom of the IRR-1 spent fuel pool (left). Data acquisition equipment (right).*

In 2022, 32 standard FAs and 3 control FAs were measured (some overlapping with the 2018 measurements), providing a close to comprehensive survey of the IRR-1 fuel inventory. A summary of all measurements performed is provided in Table A-1 of Appendix A.

## 2.2   Gamma-ray Dose Investigation

The gamma-ray dose from spent FAs is very high (on the order of hundreds of krad per hour on contact) within the first weeks after removal from a reactor. At these dose rates, pileup pulses from high-frequency gamma-ray interactions in the AEFC's $^3$He tubes can be incorrectly counted as neutrons. This results in systematic bias in the neutron measurements [9]. However, because of the gamma-ray sources' decay characteristics, the dose can decrease substantially in a matter of weeks after an irradiation. When considering the decay heat of a reactor, contributions from past irradiations must be accounted for as well. Figure 2 shows a representative calculation of the decay heat in the IRR-1, taking into account 2 months of past irradiations. This figure shows that for a campaign taking place after 2 months of cooldown (~$10^3$ hours), the last 2 months of irradiations will account for most of the decay heat. For the present campaign, the background decay heat is calculated by using [10] and summing the results from all past documented irradiations.

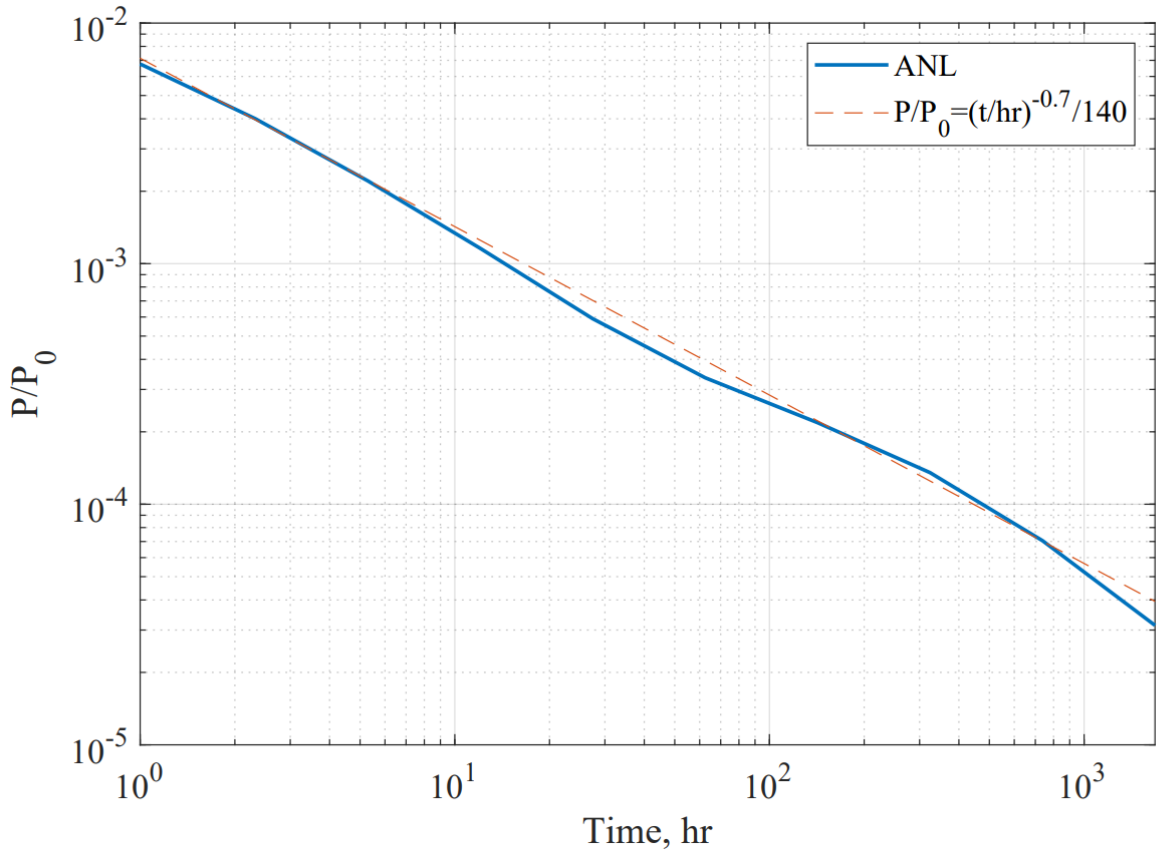

*Figure 2. Decay heat relative to irradiation power, resulting from 2 months of irradiations in the IRR-1, as a function of time. A power-law curve provides a simple estimate for the detailed calculations. Calculated using [11].*

The objectives of the gamma-ray dose investigation were to determine whether short-cooled assemblies could be measured with the AEFC at all, and, if so, how the accuracy of the measurements would be affected by the resulting decrease in the detectors' efficiency. By lowering the operating HV (and keeping the discriminator threshold of the tubes' preamplifiers constant), fewer gamma-originated pileup pulses are recorded. However, this also results in a loss in efficiency for counting true neutron interactions. As described in [2], 1620 V was the chosen operating HV for the 2018 campaign because it was the last stable HV point before gamma-ray pileup overwhelmed the detectors. The minimal cooling time for assemblies in the 2018 campaign was 3 months. In contrast, in 2022, initial measurements began only 5 weeks after the FAs were last irradiated in the core. The gamma pileup observed was, therefore, significantly higher, and required the operating voltage to be decreased to 1560 V in order to filter them out of the measurements. The calculated efficiency in the fuel measurements at 1560 V as compared to 1680 V was 75% in Singles and 53% in Doubles. The loss in efficiency decreases the Doubles count rates, thus increasing their statistical uncertainty, which already is the dominant source of uncertainty in fissile mass measurements using Doubles. However, even with the decreased HV, 2022 assays showed that short-cooled assemblies can still be measured by the AEFC with reasonable uncertainties.

To examine the effect of the gamma dose rate, we repeated the measurement of a high decay heat FA over time on a weekly basis. In each of the weekly measurements, an HV scan was generated with the FA in the center position of the AEFC fuel funnel and an active interrogation source in place. This allowed us to observe the gradual shift of the gamma pileup interference to the higher voltages of the HV plateau over time, as shown in Figure 3.

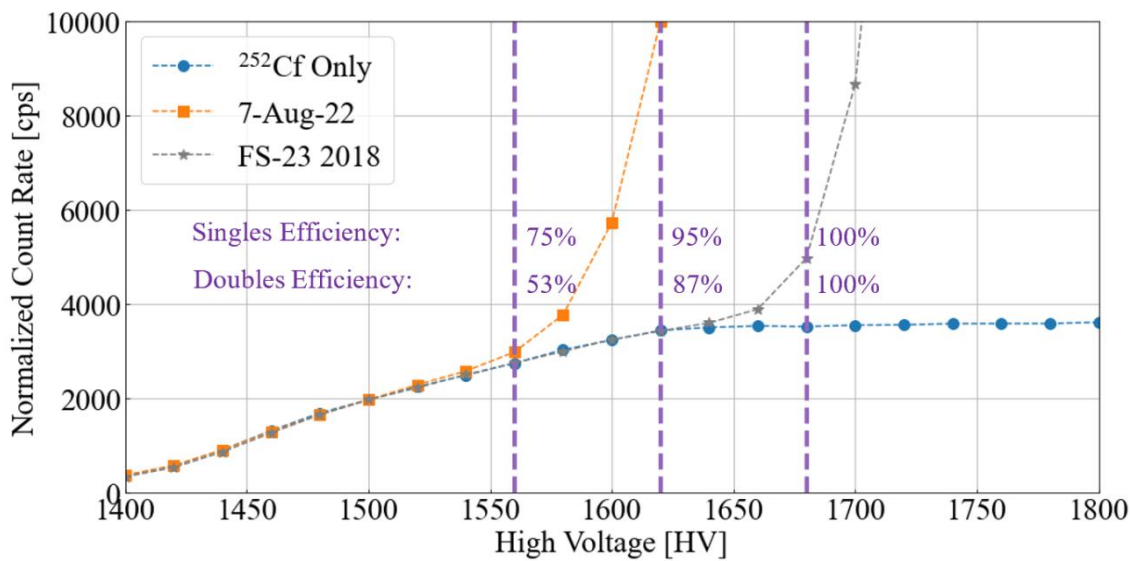

*Figure 3. Measured HV scans of FS-14 generated for different cooling times after last in-core irradiation. As the gamma flux decreases, the pileup threshold voltage increases, and the operating HV may be increased, improving the Doubles count efficiency (indicated by vertical lines).*

For the 2022 campaign, we chose FS-38 to set the operating HV because simulations predicted that it would give the highest gamma-ray dose rate at the start of the campaign and therefore would be the worst-case scenario for all FAs measured. The operating HV was set to 1560 V to minimize the impact on statistical uncertainty in measured Doubles rates from the competing effects of efficiency loss and gamma-ray pileup, as described in Section 2.3. Additionally, the campaign was designed to measure FAs with higher predicted dose at the end of the campaign to provide those FAs additional cooling time.

## 2.3 Gamma-ray Uncertainty Analysis

Following the 2018 campaign, an uncertainty quantification analysis was performed. Several different sources of uncertainty were considered and quantified. The total measurement uncertainty was found to be 3–6% in the residual $^{235}$U mass [2]. We found that in 2018, counting statistics was, by a considerable margin, the dominant source of uncertainty for the Doubles-rate-based fissile mass estimate. The following paragraphs describe the uncertainty contribution from operating at a lower HV to accommodate the higher gamma dose rate.

Pileup pulses from gamma rays are not expected to be correlated with true neutron counts and therefore should not impact the expected value of measured Doubles. However, pileup pulses contribute to the overall count rate and impact uncertainty in the Doubles calculation through the increase in accidental coincidences. We quantify this impact and present a procedure to optimize HV to minimize the Doubles uncertainty as follows.

When gamma-ray pulse pileup results in signals exceeding the $^3$He discriminator's threshold, additional gamma-generated Singles ($S_\gamma$) are counted per second in addition to the Singles rate from neutron interactions ($S_n$). The overall measured Singles ($S$) rate is

$$S = S_n + S_\gamma. \tag{1}$$

The pileup pulses are not correlated; however, the increased Singles rate results in a higher rate of accidental coincidences ($A$), given by

$$A = GS^2, \tag{2}$$

where $G$ is the coincidence gate width in seconds. This has an impact on the measured Doubles rate uncertainty because the Doubles rate ($D$) is measured indirectly and is calculated as the difference between a coincidence rate including both Doubles and accidentals and a coincidence gate counting only accidentals:

$$D = (D + A) - A. \tag{3}$$

Although there is a correlation between the two measured coincidence rates, the uncertainty of the overall Doubles rate ($\sigma_D$) can be approximated as

$$\sigma_D = \frac{\sqrt{D + 2A}}{\sqrt{t}}, \tag{4}$$

where $t$ is the measurement time in seconds.

Substituting Equation 1 gives

$$\sigma_D = \frac{\sqrt{D + 2G(S_n + S_\gamma)^2}}{\sqrt{t}}. \tag{5}$$

Thus, an increase in pileup pulses results in an increase in Doubles uncertainty.

To mitigate the impact of pileup pulses on the measured Doubles rate, the ³He tube can be operated at a lower HV, reducing the rate of pileup pulses that exceed the discriminator's threshold ($S_{\gamma,V}$ - all variables with subscript "V" change with HV). Note that it is impossible to change the discriminator threshold without accessing the AEFC's internal components, but adjusting the HV is still possible. However, a reduced HV results in lower neutron detection relative efficiency ($\varepsilon_V$). This further results in a lower measured Doubles rate relative to the Doubles rate at a fixed reference operating HV ($D_{V_0}$):

$$D_V = \varepsilon_V^2 D_{V_0}. \tag{6}$$

Accounting for the reduced efficiency, the neutron Singles rate is reduced as well, so the active fuel count becomes $S_V = S_{\gamma,V} + \varepsilon_V S_{n,V_0}$ ($S_{n,V_0}$ being the true neutron Singles rate at the reference operating HV, $V_0$), and the Doubles uncertainty at a lower HV ($\sigma_{D,V}$) is

$$\sigma_{D,V} = \frac{\sqrt{\varepsilon_V^2 D_{V_0} + 2GS_V^2}}{\sqrt{t}}. \tag{7}$$

Ultimately, the relative uncertainty in Doubles impacts overall measurement uncertainty:

$$\frac{\sigma_{D,V}}{D_V} = \frac{\sqrt{\varepsilon_V^2 D_{V_0} + 2GS_V^2}}{\varepsilon_V^2 D_{V_0} \sqrt{t}}. \tag{8}$$

Recognizing that the active reference HV plateau measurement $R_V$ (with a ²⁵²Cf interrogation source only) provides the efficiency by $R_V = \varepsilon_V R_{V_0}$

$$\frac{\sigma_{D,V}}{D_V} = \frac{\sqrt{1 + \frac{2GR_{V_0}^2}{D_{V_0}}\left(\frac{S_V}{R_V}\right)^2}}{\varepsilon_V \sqrt{D_{V_0} t}}. \tag{9}$$

Only $\varepsilon_V$ and $(S_V/R_V)$ vary with HV. Their relationship to HV can be determined from HV plateau measurements (such as those presented in Fig. 3). Typically, the accidentals are much higher than the real Doubles ($GS_V^2 \gg D_{V_0}$), so the relative uncertainty voltage curve shape is independent on the actual Doubles and Singles rate. Looking for the voltage of minimum uncertainty (zero voltage derivative) obtains the condition:

$$\frac{d}{d\varepsilon_V}\left(\frac{S_V}{R_V}\right) = \frac{1}{\varepsilon_V}\left(\frac{S_V}{R_V}\right). \tag{10}$$

HV plateau measurements of $^{252}$Cf ($R_V$) and FS-14 ($S_V$), measured on August 7[th], 2022, are shown in Figure 4. The efficiency-normalized count rate is the count rates normalized to the reference $^{252}$Cf count rate at 1620 V ($R_{V_0}$), while the $^{252}$Cf and FS-14 count rates were normalized so that they coincide at 1500 V. From 1400 V through 1500 V, the curves show good agreement with each other. However, at higher HV values, pile-up pulses from the high gamma dose result in increasing extra counts for the FS-14 measurement. The ratio between these curves as a function of HV and is plotted in Figure 5.

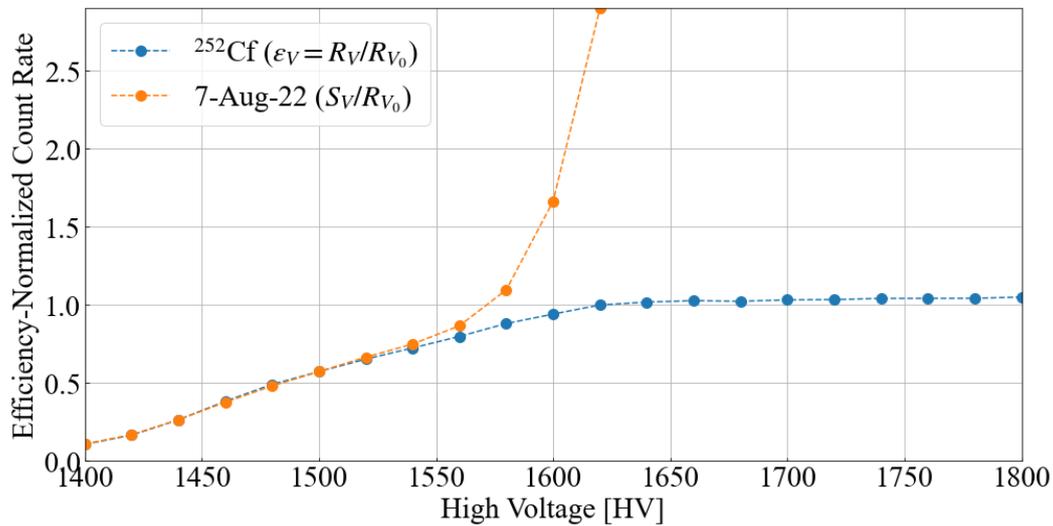

*Figure 4. HV plateau measurements of $^{252}$Cf and FS-14 measured on August 7, 2022.*

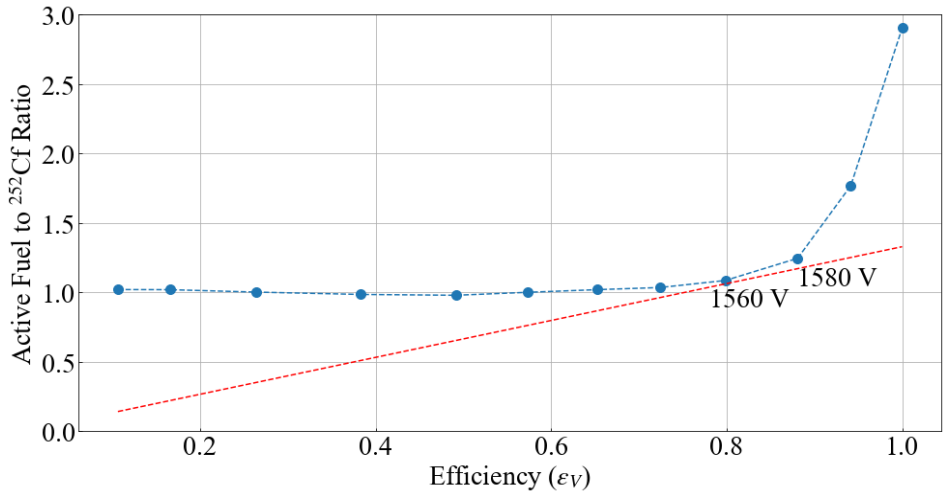

*Figure 5. Active fuel (with gamma) to active reference (interrogation source without gamma) count rate, versus relative efficiency ($\varepsilon_V$). Plotting the tangent line at HV of about 1560 V (labelled) in a red dotted line shows that is intercepts the origin*

Equation 10 has the graphical interpretation of finding the voltage where, when plotting the count rate ratio of a FA (with gamma) to Cf (no gamma) versus the efficiency, the tangent of the curve intercepts the origin, as seen in Fig. 5.

The relative Doubles uncertainty as a function of HV, as calculated using Equation 9, is plotted in Figure 6. As predicted by the approximation in Fig. 5, the uncertainty is minimized at 1560 V. Therefore, 1560 V was chosen as the operating HV throughout the 2022 campaign.

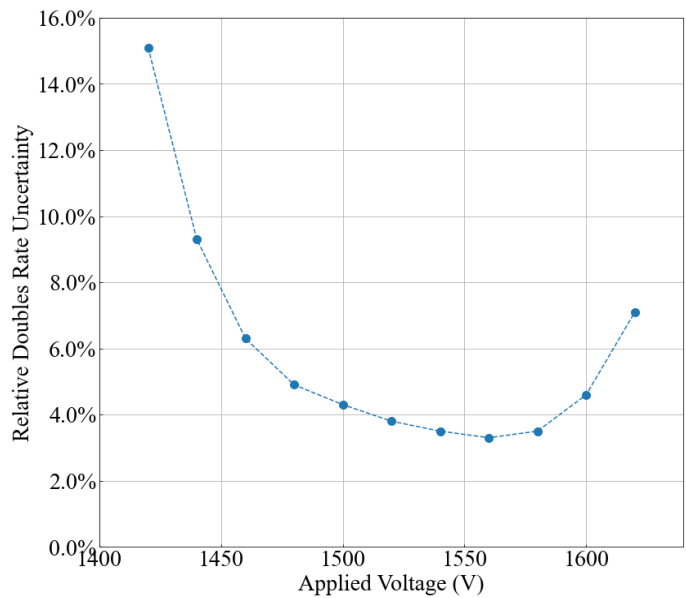

*Figure 6. Relative Doubles uncertainty as a function of HV.*

Additional variations in Singles rates beyond expected uncertainty were observed during the campaign. We hypothesize that these variations are attributable to operating the AEFC off the HV plateau such that its efficiency had a higher sensitivity to small fluctuations in HV. Section 2.4 discusses a method to compensate for these variations.

## 2.4 Normalizing Measurements to the Calibration Curve

The calibration curves developed in 2018 convert the net active neutron count rate, calculated by subtracting passive and active background rates from active fuel rates ($r_A$), to the residual fissile mass in an FA. The net active rate should be proportional to the rate of induced fission in the residual fissile mass in an FA, and it is calculated separately for Singles and Doubles so it can be applied to their respective calibration curves. The passive background rate ($r_P$) is the neutron count rate attributable to spontaneously fissioning nuclides present in the used fuel; it is determined by taking an AEFC measurement of an FA without an interrogation source present. The active background is the neutron count rate directly attributable to the interrogation source ($r_D$), and it is determined by positioning a dummy FA in the AEFC fuel funnel and measuring the neutron count rate while the interrogation source is present.

Each FA analyzed during the 2018 and 2022 campaigns was measured at three axial positions within the AEFC to account for the axial distribution of residual fissile material in the fuel plates. The results of the measurements taken at these three positions were summed to create a single value corresponding to the total residual fissile mass of each FA. The net active rate ($r_{net}$) for a given FA is calculated as shown in Equation 9.

$$r_{net} = \sum_{i=1}^{3} r_{A,i} - r_{P,i} - r_{D,i} \qquad (9)$$

To apply the 2018 calibration curves to the net active Singles and Doubles measurements taken during the 2022 campaign, the 2022 measurements were normalized to account for changes in operating conditions [12]. The 2018 and 2022 measurement conditions differed in two primary respects: the operating HV of the AEFC, and the $^{252}$Cf source strength.

As described in previous sections, in order that the large gamma dose from irradiated fuel not to interfere with the $^{3}$He tubes' neutron counting, the AEFC was operated at an HV of 1560 V during the 2022 measurement campaign. This resulted in the detection efficiency loss described in Section 2.2., which reduced the measured net active neutron count rates for a given assembly. Additionally, the 2022 measurement campaign used a stronger $^{252}$Cf interrogation source than was used in 2018. The stronger interrogation source increased the induced fission rate, increasing the net active rate measured for a given assembly. To account for the combined and competing effects of reduced operating HV and increased $^{252}$Cf source strength, the 2022 measurements were normalized to the operating conditions of 2018 using a series of reference $^{252}$Cf-only measurements.

The normalization was accomplished by calculating a dynamic Singles and Doubles correction factor for each assembly's three-point scan. This correction factor was a ratio of a 2018 $^{252}$Cf reference measurement to one of a series of $^{252}$Cf measurements performed during the 2022 measurement campaign. The 2018 reference measurement was performed in January 2018 at 1620 V using a $^{252}$Cf source with an approximate neutron emission rate of 135,600 n/s. By directly comparing this reference measurement with $^{252}$Cf measurements taken at 1560 V using the 3.5 times stronger source, the 2022 measurements were each normalized using a dynamic correction factor that accounted for differences in both the $^{252}$Cf source strength and the detection efficiency loss. A series of $^{252}$Cf measurements at 1560 V were collected throughout the 2022 campaign to monitor the stability of the AEFC efficiency and these were ultimately used to calculate the correction factors. Figure 7 shows the time distribution of the $^{252}$Cf Singles measurements used in the correction factor calculations.

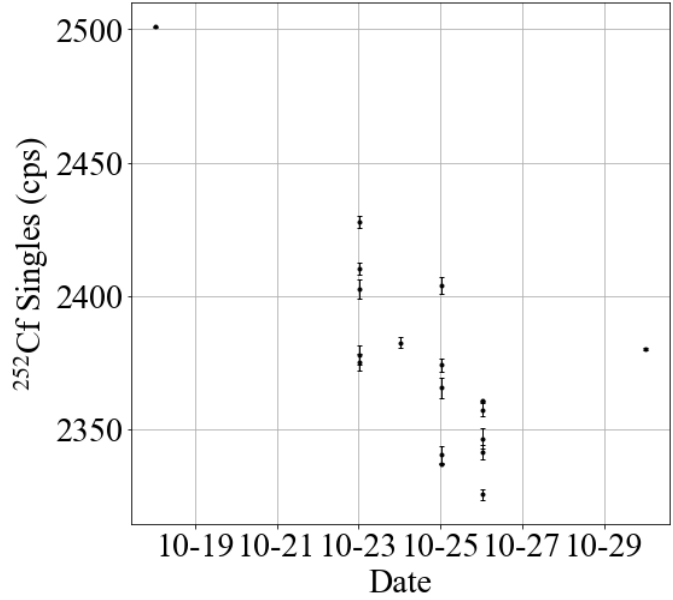

*Figure 7. $^{252}$Cf Singles rate measurements taken throughout the campaign. Error bars represent 1σ statistical uncertainty in the measured count rates.*

The Singles rates from these measurements were found to fluctuate far beyond expected statistical uncertainties, indicating the presence of some instability in the AEFC detection efficiency during the campaign. To account for these efficiency fluctuations, the Singles and Doubles correction factors for each assembly were calculated using the results of the $^{252}$Cf Singles measurement that took place closest in time to its three-point scan.

Many of the $^{252}$Cf measurements took place over a count time that resulted in satisfactory counting statistics for the Singles measurement but in poor statistics for the Doubles rate. Although each assembly's Singles correction factor ($C_S$) could be calculated using a simple ratio of the 2018 $^{252}$Cf reference measurement ($S_{ref,2018}$) to its proximal 2022 $^{252}$Cf measurement ($S_{ref}$), as shown in Equation 10, a different approach was required for each assembly's Doubles correction factor due to the poor counting statistics present in many of the 2022 $^{252}$Cf Doubles rates. A linear relationship was established between $^{252}$Cf Singles and Doubles rates measured at 1560 V using four $^{252}$Cf measurements taken with sufficient count times to reduce the uncertainty in the Doubles rates to the order of 1% (refer to Figure A-1).

The Doubles correction factor for each assembly ($C_D$) was calculated as shown in Equation 11 using a ratio of the 2018 $^{252}$Cf Doubles rate ($D_{ref,2018}$) and the linearly regressed 2022 $^{252}$Cf Singles rate, where *p* and *q* represent the coefficients of the linear transformation (refer to Figure A-1).

$$C_S = \frac{S_{ref,2018}}{S_{ref}} \quad (10)$$

$$C_D = \frac{D_{ref,2018}}{pS_{ref} + q} \quad (11)$$

The 2022 data were normalized by correcting the sum of the differences between the active and passive measurements at each position for each FA using the correction factors defined above (Equation 12). The net active rate for each assembly was then calculated by subtracting the active background from 2018 ($r_{dummy,2018}$) from the normalized 2022 data ($r_{norm}$) (Equation 13). Finally, the 2018 calibration curves [2], with coefficients *a* and *b*, were applied to the net active rates from the 2022 measurements to predict the residual fissile mass ($m$) in the IRR-1 used FAs (Equation 14).

$$r_{norm} = C \times \left( \sum_{i=1}^{3} r_{A,i} - r_{P,i} \right) \qquad (12)$$

$$r_{net} = r_{norm} - r_{dummy,2018} \qquad (13)$$

$$m = a r_{net} + b r_{net}^2 \qquad (14)$$

## 2.5 Measured Residual $^{235}$U Mass Uncertainty Analysis

Each of the measured residual $^{235}$U mass values has an associated uncertainty, which represents a combination of statistical measurement uncertainty in the AEFC neutron Singles and Doubles rates, systematic uncertainty associated with the parameters of the calibration curves, and additional random errors that are a part of any regression analysis. The uncertainty for each $^{235}$U mass estimate was determined by first calculating its calibration uncertainty ($\sigma_{cal}$), that is, the uncertainty attributable to its calibration curve. Then, the measurement uncertainty ($\sigma_{stat}$) was calculated by propagating the statistical uncertainty in each AEFC measurement to the net active neutron Doubles and Singles rates ($\sigma_x$, see section 2.3) and then to the residual $^{235}$U mass estimates. Calibration and measurement uncertainties for each fuel assembly $^{235}$U mass estimate were treated as independent, so total uncertainty in the $^{235}$U mass estimate was calculated by summing the calibration and measurement uncertainties in quadrature. The procedure for calculating the total uncertainty for an assembly $^{235}$U mass estimate is summarized in Equations 15–17. Uncertainty in each $^{235}$U mass estimate that was attributable to the calibration curve was calculated using Equation 15. The standard errors ($\sigma_a$ and $\sigma_b$) and covariance ($cov(a,b)$) of the linear ($a$) and quadratic ($b$) coefficients in the calibration curves ($f$) were derived from the covariance matrices of the points used to fit the calibration curves. Calibration uncertainty in each assembly's fissile mass estimate is a function of the net active count rate ($x$) measured from that assembly.

$$\sigma_{cal} = \sqrt{\left(\frac{\partial f}{\partial a}\right)^2 \sigma_a^2 + \left(\frac{\partial f}{\partial b}\right)^2 \sigma_b^2 + 2 \frac{\partial f}{\partial a} \frac{\partial f}{\partial b} cov(a,b)}$$
$$= \sqrt{x^2 \sigma_a^2 + x^4 \sigma_b^2 + 2 x^3 cov(a,b)} \qquad (15)$$

The statistical measurement uncertainty for each assembly $^{235}$U mass estimate was calculated using Equation 16.

$$\sigma_{stat} = \sqrt{\left(\frac{\partial f}{\partial x}\right)^2 \sigma_x^2} = \sqrt{(a + 2bx)^2 \sigma_x^2} \qquad (16)$$

Finally, to determine total uncertainty in each fissile mass estimate ($\sigma_{tot}$), statistical and calibration uncertainties were summed in quadrature, as described in Equation 17.

$$\sigma_{tot} = \sqrt{\sigma_{cal}^2 + \sigma_{stat}^2} \qquad (17)$$

## 3. Results

After the 2022 measurement campaign, the calibration curves developed during the 2018 measurement campaign [2] were used to estimate the residual $^{235}$U mass in each of the 32 standard FAs using their net active neutron Doubles and Singles rates. These measured residual $^{235}$U mass values were compared to calculated residual $^{235}$U mass values estimated from detailed IRR-1 core depletion simulations. The comprehensive campaign measurement results are presented in Figures 8 and 9. The solid blue lines in Figures 8 and 9 represent agreement between the measured and calculated residual $^{235}$U mass values, and the black plotted points indicate the measurement and core depletion mass estimates for each assembly. The error bars represent a 1σ total uncertainty in the measured residual $^{235}$U mass that accounts for both statistical and systematic uncertainties. The details of the total uncertainty calculations are described in Section 4.2.

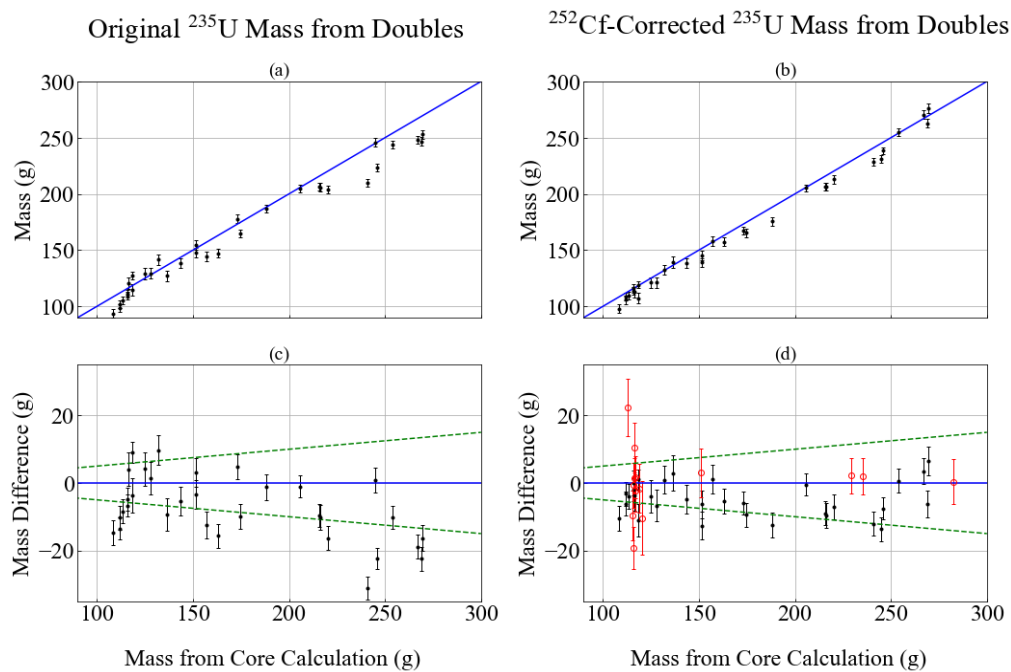

*Figure 8. Plots (a) and (b) show original (static) and $^{252}$Cf-corrected (dynamic) $^{235}$U mass estimates from measured net active Doubles rates versus expected $^{235}$U mass values from core calculations. Error bars represent 1σ total uncertainty in the measured $^{235}$U estimates. Plots (c) and (d) show the deviation of measured $^{235}$U mass from depletion calculations. The dashed lines represent a 5% deviation from the expected value. The red dots in (d) show the results from the 2018 campaign.*

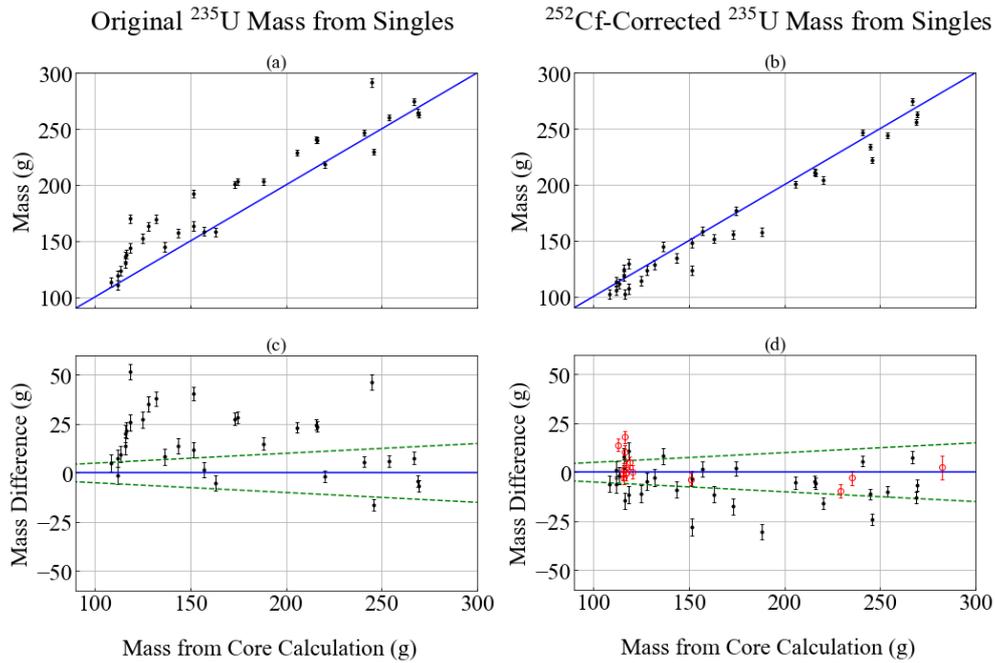

*Figure 9. Plots (a) and (b) show original (static) and $^{252}$Cf-corrected (dynamic) $^{235}$U mass estimates from measured net active Singles rates versus expected $^{235}$U mass values from core calculations. Error bars represent 1σ total uncertainty in the measured $^{235}$U estimates. Plots (c) and (d) show the deviation of measured $^{235}$U mass from depletion calculations. The dashed lines represent a 5% deviation from the expected value. The red dots in (d) show the results from the 2018 campaign.*

## 4. Discussion

### 4.1 Measured Residual $^{235}$U Mass Results

Prior to the development of the correction factor method outlined in Section 2.4, net active neutron Doubles and Singles rates were initially calculated using a correction factor derived from the known interrogation source strength and a static, estimated efficiency correction (plots (a) and (c) of Figures 8 and 9). The measured residual $^{235}$U mass values calculated using this static correction factor demonstrated much greater deviation than expected from the $^{235}$U mass values predicted by core calculations, particularly for the Singles. After performing the static correction factor net active neutron count rate residual $^{235}$U mass estimates, it was observed that the deviation of the measured Singles residual $^{235}$U mass values from the calculated $^{235}$U mass was much greater than was observed in the previous measurement campaign. This was partially attributed to the observed variation in $^{252}$Cf Singles rates beyond statistical uncertainties throughout the 2022 campaign, as shown in Figure 7 in Section 2.4. As a result, the dynamic $^{252}$Cf correction factor method outlined in Section 2.4 was developed and applied to the net active neutron count rate calculations; then, the calibration curves were applied to the $^{252}$Cf-corrected net active Doubles and Singles rates to predict the residual $^{235}$U mass (plots (b) and (d) of Figures 8 and 9).

Comparing the static and dynamic plots in Figures 8 and 9 shows that the $^{252}$Cf-corrected measured $^{235}$U mass values perform better than the original measured $^{235}$U mass values with respect to the expected residual $^{235}$U mass values from the core depletion calculation. The RMSE for the net active Doubles residual $^{235}$U mass calibration curve improves from 6.74% for the original measured $^{235}$U mass to 4.37% for the $^{252}$Cf-corrected $^{235}$U mass. Using the net active Singles calibration curve, the RMSE decreases from 15.48% for the original measured

$^{235}$U mass to 7.14% for the $^{252}$Cf-corrected $^{235}$U mass. The residual $^{235}$U mass values from the core calculation and the $^{252}$Cf-corrected measured residual $^{235}$U mass values using both the net active Doubles and Singles are tabulated in Appendix A Table A-2.

Table A-2 also records the calculated contact dose rate of each assembly, as described in Section 2.2. Deviations of measured $^{235}$U mass estimates from expected $^{235}$U mass values from core depletion calculations were not found to vary as a function of assembly dose rate. This result suggests that the modifications made to the AEFC operational procedure to mitigate the effects of a large gamma dose did not introduce additional bias into the $^{235}$U mass estimates from the neutron Doubles and Singles rates. By taking steps to account for the effects of gamma pileup, the AEFC can be effectively used to assay spent-fuel assemblies with short cooling times since their last irradiation - it tolerates contact dose rates in the order of 100 krad/h.

## 4.2 Measured Residual $^{235}$U Mass Uncertainty Analysis

The uncertainties for the measured $^{235}$U mass estimates are tabulated in Table A-2. The residuals plots in Figure A-2 show that the magnitude of measured $^{235}$U mass deviations is greater than the magnitude of the total uncertainty in measured $^{235}$U mass, which indicates the presence of some additional error or uncertainty in the methodology that is not captured by statistical uncertainty in neutron count rate measurements or by systematic uncertainty in calibration curve parameters. Note that the total uncertainties do not directly account for uncertainties in core-calculated $^{235}$U mass values; however, uncertainties in the core calculation are on the order of 1% and are expected to be dwarfed by the systematic and statistical uncertainty of measured $^{235}$U mass values.

The reported total uncertainties for the 2022 net active Doubles $^{235}$U mass estimates are smaller in magnitude than those reported for the $^{235}$U mass estimates performed in 2018. During the 2022 measurement campaign, the AEFC measurement procedure was modified to introduce a longer count time of 8 minutes for each measurement type and position. Increasing the count time and interrogation source strength resulted in an increased count rate and a reduced counting statistical uncertainty for the neutron Doubles rates. In 2018, when each AEFC measurement took place over a period of 5 minutes, the average relative uncertainty in net active neutron Doubles rates was 4.95%, and statistical uncertainty represented the dominant term in the total uncertainty for net active Doubles $^{235}$U mass estimate. In 2022, when the count times were increased to 480 seconds, the average relative uncertainty in net active neutron Doubles rates was 1.84%.

During the uncertainty analysis, it was observed that the regression analysis used to generate the calibration curves in 2018 treated uncertainties in sample observations as purely statistical. That is, any error between the calibration curve and the observed samples was attributed to uncertainty in the measured neutron Singles and Doubles rates. This treatment ignored "goodness-of-fit" for the calibration curves when calculating coefficient uncertainties. Analysis in this work considered the possibility of additional random sources of uncertainty in the calibration curves fits. This new statistical treatment increased the uncertainties of the Singles calibration curve coefficients and reduced the uncertainties of the Doubles calibration curve coefficients. These changes in calibration curve coefficient uncertainties were ultimately propagated to the reported total uncertainty for each $^{235}$U mass estimate.

## 5. Conclusion

This paper summarizes the results of an AEFC measurement campaign that took place at IRR-1 in 2022. The campaign measured 32 standard FAs and used calibration curves developed in 2018 to estimate the residual $^{235}$U mass in the assemblies from AEFC measurements. This is the most complete inventory measured by the AEFC at a reactor, representing a significant majority of the irradiated fuel assemblies in the pool. The calibration curve mass estimates demonstrated close agreement with expected $^{235}$U mass values. This agreement indicates the robustness of the AEFC to variations in operational conditions, provided steps are taken to correct for those variations. In this campaign, the AEFC operating procedure was modified to use a lower HV to mitigate the effects of gamma pileup, allowing measurement of fuel assemblies with a contact dose rate of around 100 krad/h (Table A-2). This modification reduced the efficiency of the AEFC measurements and potentially introduced some instability to that efficiency. By introducing a dynamic $^{252}$Cf correction factor, it was possible to adequately correct the 2022 measurements for the efficiency variations and apply the 2018 calibration curves to estimate residual $^{235}$U mass. It is also possible that the efficiency instability could be mitigated by future upgrades to the AEFC electronics package.

The analysis of the 2022 measurement campaign results also identified opportunities for further study. In addition to the 32 standard FAs measured in 2022, 3 FCs with neutron-absorbing control blades were measured using the AEFC. Attempting to apply the 2018 calibration curves to net active neutron doubles and singles rates from these assemblies resulted in $^{235}$U mass estimates well below expected values because of the absorption of induced-fission neutrons into control blades. A measurement campaign took place at LANL in 2023 to explore the effects of such control elements on AEFC results [13]. Future work will seek to correct the 2022 IRR-1 FC measurements for the presence of Ag-In-Cd control elements and estimate the residual $^{235}$U mass in the FCs.

Deviations between residual $^{235}$U mass estimates and expected $^{235}$U mass values from core depletion calculations indicate that there could be some linear bias included in net active neutron doubles and singles rate calculations that propagates to $^{235}$U mass estimates. Although there is no evident trend in the residuals plot (Figure A-2), the residuals do not have a mean of zero, and the calibration curves tend to underpredict $^{235}$U mass when compared to core calculation estimates. It is possible that this shift in the mean is attributable to subtraction of the active background, a static value, during net active count rate calculations. Exploring other methods for the treatment of the active background when calculating net active neutron Doubles and Singles rates could also be of interest.

## CRediT

**G. Long**: Formal analysis, Investigation, Writing – original draft, Writing – review and editing. **G. Gabrieli**: Formal analysis, Investigation, Writing – original draft, Writing – review and editing. **I. Levy**: Formal analysis, Investigation, Writing – original draft, Writing – review and editing. **A. Pesach:** Formal analysis, Investigation, Writing – original draft. **I. Zilberman**: Investigation. **K. Ben-Meir**: Investigation. **O. Ozeri**: Investigation. **L. Zilberfarb**: Investigation. **O. Rivin**: Investigation. **A. Krakovich**: Investigation. **I. Martinez**: Investigation. **C. Rael**: Investigation. **M. Watson**: Investigation. **A. Trahan**: Investigation. **M. L. Ruch**: Formal analysis, Investigation, Writing – original draft, Writing – review and editing. **U. Steinitz:** Formal analysis, Investigation, Writing – original draft, Writing – review and editing.

## Acknowledgements

This work was supported by the Office of International Nuclear Safeguards within the U.S. Department of Energy's National Nuclear Security Administration and by the Israel Atomic Energy Commission.

# Appendix A. Measurement Summaries

*Table A-1: Summary of measurements performed in the 2018 and 2022 campaigns, spanning most of the inventory [14].*

|  | 2018 | | | | 2022 | | | |
|---|---|---|---|---|---|---|---|---|
|  | 3-Point Scan | 15-Point Scan | HV Plateau | Rotations | 3-Point Scan | 11-Point Scan | HV Plateau | Rotations |
| FS-1 | X | | | | | | | |
| FS-2 | X | X | | | X | | | X |
| FS-3 | X | | | | | | | |
| FS-4 | | | | | | | | |
| FS-5 | X | X | | | | | | |
| FS-6 | X | | | | | | | |
| FS-7 | | | | | X | | | |
| FS-8 | | | | | | | | |
| FS-9 | X | | | | | | | |
| FS-10 | | | | | X | | | |
| FS-11 | X | | | | X | | | |
| FS-12 | | | | | X | | | |
| FS-13 | | | | | X | | | |
| FS-14 | | | | | | | | X |
| FS-15 | | | | | X | X | | |
| FS-16 | | | | | | | | |
| FS-17 | | | | | X | | | |
| FS-18 | X | X | | | X | | | |
| FS-19 | | | | | X | | | X |
| FS-20 | X | | | | X | | | |
| FS-21 | | | | | | | | |
| FS-22 | | | | | X | | | X |
| FS-23 | X | | X | X | X | | | |
| FS-24 | | | | | X | | | X |
| FS-25 | | | | | X | X | | |
| FS-26 | | | | | X | | | |
| FS-27 | | | | | X | | | X |
| FS-28 | | | | | X | | | |
| FS-29 | | | | | X | | | |
| FS-30 | | | | | X | | | |

| | | | | | | | |
|---|---|---|---|---|---|---|---|
| FS-31 | | | | X | | | X |
| FS-32 | X | | X | X | | | |
| FS-33 | | | | X | | | |
| FS-34 | | | | X | | | |
| FS-35 | | | | X | X | | |
| FS-36 | | | | X | | | |
| FS-37 | | | | X | | | |
| FS-38 | | | | X | | | |
| FS-39 | | | | X | | | |
| FS-40 | X | | X | X | | | X |
| FS-41 | | | | X | X | | |
| FS-42 | | | | X | | | X |
| FS-43 | | | | | | | |
| FS-44 | | | | | | | |
| FS-45 | X | | | | | | |
| FS-D | X | | | X | | | |
| FC-1 | X | | | X | X | | |
| FC-4 | | | | X | | | X |
| FC-5 | | | | X | X | | |

Table A-2. Results of residual $^{235}$U mass determination using core calculation, net active Doubles calibration curve, and net active Singles calibration curve with associated 1σ uncertainties, spanning most of the inventory

| Assembly | Contact Dose Rate (krad/h) | Residual $^{235}$U from core calc (g) | Residual $^{235}$U from Doubles (g) | σ (Residual $^{235}$U from Doubles (g)) | Residual $^{235}$U from Singles (g) | σ (Residual $^{235}$U from Singles (g)) |
|---|---|---|---|---|---|---|
| FS-13 | 4.73 | 108.4 | 97.9 | 3.7 | 102.5 | 4.3 |
| FS-10 | 4.50 | 112.0 | 108.8 | 3.6 | 113.2 | 4.3 |
| FS-07 | 3.93 | 112.0 | 105.8 | 3.6 | 105.7 | 4.3 |
| FS-15 | 6.05 | 113.5 | 109.7 | 3.7 | 111.6 | 4.3 |
| FS-16 | 10.13 | 116.0 | 114.1 | 3.4 | 124.0 | 4.3 |
| FS-12 | 3.92 | 116.1 | 116.1 | 3.6 | 118.1 | 4.3 |
| FS-20 | 51.14 | 116.6 | 113.0 | 4.7 | 102.0 | 4.3 |
| FS-02 | 0.49 | 118.2 | 119.3 | 3.0 | 129.1 | 4.3 |
| FS-18 | 37.36 | 118.3 | 107.3 | 4.9 | 106.9 | 4.3 |
| FS-22 | 84.10 | 125.0 | 121.1 | 4.7 | 113.9 | 4.3 |
| FS-17 | 70.07 | 128.0 | 121.3 | 4.7 | 123.3 | 4.3 |
| FS-19 | 65.31 | 131.8 | 132.7 | 4.1 | 129.0 | 4.3 |
| FS-23 | 90.06 | 136.5 | 139.4 | 5.2 | 144.8 | 4.1 |
| FS-25 | 78.34 | 143.4 | 138.5 | 4.3 | 134.5 | 4.2 |
| FS-24 | 89.97 | 151.5 | 145.2 | 4.1 | 148.2 | 4.1 |
| FS-27 | 90.54 | 151.7 | 139.0 | 4.0 | 123.6 | 4.3 |
| FS-26 | 93.23 | 157.0 | 158.0 | 4.2 | 158.6 | 3.9 |
| FS-30 | 97.57 | 163.0 | 157.6 | 3.7 | 151.8 | 4.0 |
| FS-28 | 56.41 | 173.0 | 167.1 | 3.5 | 155.5 | 4.0 |
| FS-29 | 90.52 | 174.8 | 165.4 | 3.6 | 176.8 | 3.6 |
| FS-31 | 81.78 | 188.1 | 175.6 | 3.6 | 157.6 | 3.9 |
| FS-33 | 73.68 | 205.8 | 205.3 | 3.2 | 200.5 | 3.1 |
| FS-34 | 91.39 | 215.8 | 206.8 | 3.5 | 211.1 | 3.0 |
| FS-32 | 79.49 | 216.1 | 206.5 | 3.7 | 210.5 | 3.0 |
| FS-35 | 79.76 | 220.1 | 213.0 | 3.7 | 204.2 | 3.1 |
| FS-36 | 75.10 | 240.8 | 228.7 | 3.6 | 246.4 | 2.7 |
| FS-38 | 114.08 | 245.1 | 231.4 | 3.5 | 234.0 | 2.7 |
| FS-37 | 100.33 | 245.9 | 238.4 | 3.2 | 221.9 | 2.8 |
| FS-39 | 96.34 | 254.2 | 254.8 | 3.5 | 244.3 | 2.7 |
| FS-41 | 88.02 | 266.9 | 270.4 | 3.8 | 274.4 | 3.2 |
| FS-42 | 100.73 | 269.1 | 262.9 | 4.0 | 256.2 | 2.8 |
| FS-40 | 77.06 | 269.5 | 275.9 | 4.2 | 263.0 | 2.9 |

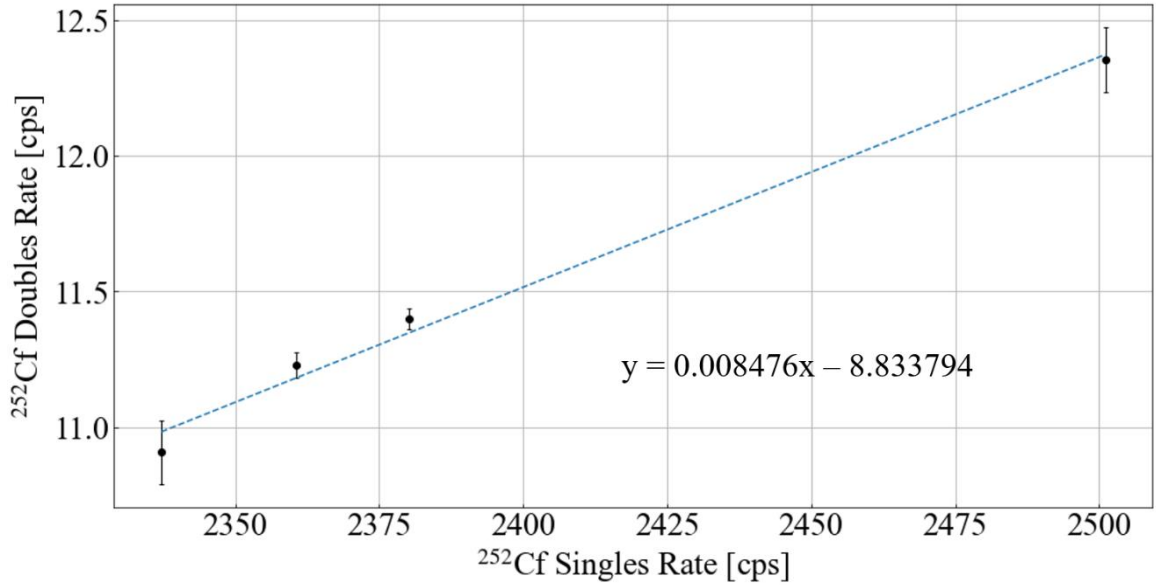

*Figure A-1. Linear relationship between $^{252}$Cf Singles and Doubles rates.*

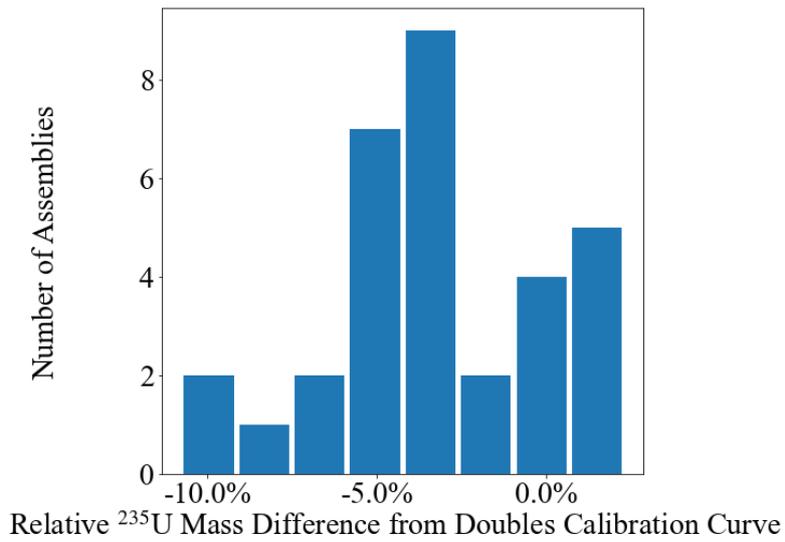

*Figure A-2. Histogram of relative measured $^{235}$U mass deviations from expected core calculation values.*